\documentclass[amssymb,aps,floats,twocolumn]{revtex4}
\usepackage{color,graphicx,pstricks}

\begin{document}
\title{ Combined Effect of Bond- and Potential-Disorder in Half-Doped Manganites }

\author{Sanjeev Kumar\footnote{Present address: Faculty of Science and Technology, University of Twente, P.O. Box 217,
7500 AE Enschede, The Netherlands.} and Arno P. Kampf }

\affiliation{~ Theoretical Physics III, Center for Electronic Correlations and Magnetism,
Institute of Physics, University of Augsburg, D-86135 Augsburg, Germany }

\begin{abstract}
We analyze the effects of both
bond- and potential-disorder in the vicinity of a first-order metal insulator transition
in a two-band model for manganites using a real-space Monte Carlo method.
Our results reveal a novel charge-ordered state coexisting with spin-glass 
behavior. We provide the basis for understanding the phase diagrams of half-doped manganites,
and contrast the effects of bond- and potential-disorder and the combination of both.

\end{abstract}
\vskip0.15cm

\maketitle

The perovskite manganites R$_{1-x}$A$_x'$MnO$_3$ (R = rare earth, A$'$ = alkaline earth) 
have received attention from the condensed matter community, largely
due to their colossal magnetoresistance (CMR) effect \cite{dagotto_book, tapan_book}.
In these materials different ordering tendencies of charge, spin, lattice, and orbital
degrees of freedom compete. Due to this complexity the understanding of the sensitive influence of 
quenched disorder has become one of the central issues in manganites' research.
Based on a variety of experiments, Tomioka and Tokura
have shown that the observed phases near half-doping
can be organized in terms of two parameters, the average radius $r_A =(1-x)r_R+xr_{A'} $
and the variance $\sigma^2$ of the ionic radii \cite {tokura_gpd}. Since $r_A$ and $\sigma^2$ control the 
single-particle bandwidth and the amount of disorder, respectively, quenched disorder is thereby identified 
as one key parameter in manganites.
This is supported via a set of experiments on the
"ordered" and "disordered" compositions of R$_{0.5}$Ba$_{0.5}$MnO$_3$,
with a sequence of decreasing $r_A$ as R ranges from La to Y \cite {o_d, akahoshi}.
The groundstate for the "ordered" materials changes from 
a ferromagnetic metal (FM-M) to an A-type antiferromagnetic (AF) insulator near R=Pr, eventually followed by a
charge and orbitally ordered insulator with ferromagnetically aligned spins along zig-zag chains, which is commonly
referred to as the CE phase.
In the "disordered" materials the AF phase does not exist and
the Curie temperature ($T_C$) is suppressed. A spin-glass (SG) phase
is observed for R ranging from Sm to Dy. "Disordered" Nd$_{0.5}$Ba$_{0.5}$MnO$_3$ 
undergoes an insulator to metal transition near $T_C$, which is characteristic of CMR materials.
On applying an external magnetic field the spin-glass state can be driven towards a ferromagnetic metallic phase
\cite {tomioka, maignan}.

Theoretical studies of manganite-specific models at half-doping have indeed found a
variety of ordered phases and transitions for clean systems \cite{dag_x0.5, brey_x0.5}.
Among them, the first-order phase transitions are particularly interesting, since
these are assumed to split into two second-order transitions 
with an intermediate region of macroscopic phase coexistence
upon including quenched disorder \cite {dagotto_book}.
Detailed model studies to verify this assumption have only recently begun \cite{sen_PRL}. 
However, the modeling of disorder, which originates from different microscopic sources,
has remained a matter of choice.
Since the R or A$'$ ions are located away from the electronically active Mn-O$_2$ planes,
the primary effect of ionic-size mismatch is to 
modify the Mn-O-Mn bond angles \cite{attfield, fontcuberta}. A realistic model ansatz for
disorder should necessarily account for these bond angle variations.
Additionally, the random positions of differently charged R$^{3+}$ and Ba$^{2+}$ ions results in an effective on-site disorder
via Coulomb interactions. 
Therefore, both these sources of disorder are important and unavoidably occur together in doped manganites.

In this Letter, we provide a systematic analysis for the phase diagrams of
R$_{0.5}$Ba$_{0.5}$MnO$_3$ by modeling the Mn-O-Mn bond angle variations by angle dependent hopping 
parameters in a two-dimensional two-band double-exchange model
with electron-lattice and electron-electron interactions. We use the combination of a real-space
Monte Carlo method and an unrestricted Hartree-Fock scheme.
Bond-disorder alone explains the suppression of the Curie
temperature and the existence of a spin-glass phase, but its
effect on the charge and orbital ordering is weak. Therefore, a phase is formed 
with coexisting charge and orbital order and glassiness in the spin degree of freedom. 
We identify a thermally driven
metal to insulator transition, as well as a magnetic field
driven spin-glass insulator to ferromagnetic metal transition. 
We find that disorder does not lead to macroscopic phase coexistence, but rather
generates inhomogeneities on the scale of few lattice spacings.

Specifically, we choose a two-band model with quenched disorder for itinerant $e_g$ 
electrons coupled to localized $S=3/2$ $t_{2g}$ spins and to the Jahn-Teller (JT) lattice distortions.
The inter-orbital Hubbard repulsion $U'$ between the $e_g$ electrons and
the AF superexchange $J_S$ between neighboring $t_{2g}$ core
spins are also included.

Based on the results of previous analyses \cite{large-JH} we adopt the double-exchange limit
for the Hund's rule coupling, leading to the Hamiltonian:
\begin{eqnarray}
H &=& \sum_{\langle ij \rangle}^{\alpha \beta}
t^{\alpha \beta}_{ij} ~ f_{ij} ~
c^{\dagger}_{i \alpha} c^{~}_{j \beta } + \sum_i (\epsilon_i-\mu) n_i
+ J_S\sum_{\langle ij \rangle} {\bf S}_i \cdot {\bf S}_j \cr
&&
- \lambda \sum_i {\bf Q}_i {\bf \cdot} {   {\mbox {\boldmath $ \tau $}} }_i 
+ {K \over 2} \sum_i |{\bf Q}_i|^2 + U'\sum_{i} n_{i a} n_{i b} .~ ~ ~ ~ ~ ~
\end{eqnarray}
\noindent
In Eq. (1), $\alpha$, $\beta $ are summed over the two Mn-$e_g$ orbitals
$d_{x^2-y^2}$ ($a$) and $d_{3z^2-r^2}$ ($b$).
The operator $ c^{}_{i \alpha}$ ($c^{\dagger}_{i \alpha}$) annihilates (creates) an
electron at site $i$ in orbital $\alpha$ with its spin slaved along the direction of the
$t_{2g}$ spin ${\bf S}_i$.
$t_{ij}^{\alpha \beta}$ denote the hopping matrix elements between
$e_g$ orbitals on nearest-neighbor Mn ions via the oxygen 2p orbitals, and
hence depend on the Mn-O-Mn bond angle $\phi_{ij}$.
Taking into account the $pd\sigma$ contributions only, the hopping
parameters between neighboring sites i and j are given by \cite{slater_koster}:
\begin{eqnarray}
t_{ij}^{a a} &=& tcos^3(\phi_{ij}) ; ~  t_{ij}^{b b} = (t/3)cos(\phi_{ij}); \nonumber \\
t_{ij,x(y)}^{a b} &\equiv& t_{ij,x(y)}^{b a} = +(-)(t/\sqrt{3})cos^2(\phi_{ij}) .
\end{eqnarray}
Here, $x$ and $y$ denote the spatial directions on a square lattice and
$t = 3/4 (pd\sigma)^2$ is the basic energy unit.
The factors $f_{ij} = cos({ \Theta}_i/2)cos({\Theta}_j/2)
+ sin({\Theta}_i/2)sin({\Theta}_j/2)~e^{ -{\rm i}({\Phi}_i - {\Phi}_j) } $
are a consequence of projecting out fermions with spins antialigned to the
core-spin directions. ${\Theta}_i$ and ${\Phi}_i$ are the polar and azimuthal angles
determining the orientation of the $t_{2g}$ spin ${\bf S}_i$.

Bond-disorder arises from a non-uniform distribution of the bond-angles
$\phi_{ij}$; on-site disorder enters via the 
local potentials $\epsilon_i$, for which equally probable values $\pm \Delta$ are assumed.
$\lambda$ denotes the strength of the JT coupling between the distortion 
${\bf Q}_i = (Q_i^x, Q_i^z)$ and  
the orbital pseudospin
${\tau}^{\mu}_i = \sum^{\alpha \beta}_{\sigma}
c^{\dagger}_{i\alpha \sigma} 
\Gamma^{\mu}_{\alpha \beta} c^{~}_{i\beta \sigma}$,
where $\Gamma^{\mu}$ are the Pauli matrices
\cite{dagotto_book}. 
The spins are treated as classical
unit vectors, $\vert {\bf S}_i \vert =1$, and the lattice variables are
considered in the adiabatic limit. $\mu$ denotes the chemical potential and the lattice stiffness $K$ is set to $1$.

The average ionic radius $r_A$ in R$_{0.5}$Ba$_{0.5}$MnO$_3$ is modelled by the average angle $\phi_0$.
The "ordered" compounds are represented by a uniform Mn-O-Mn angle 
$\phi_{ij} \equiv \phi_0$, and for the "disordered" systems the angles $\phi_{ij}$
are selected from a binary distribution with mean $\phi_0$ and variance 
$\delta \phi$. Since the amount of disorder depends on
the difference in ionic radii of R and Ba ions, $\phi_0$ and $\delta \phi$
are not independent. This is reflected in the similar
behavior of "ordered" and "disordered" La$_{0.5}$Ba$_{0.5}$MnO$_3$, due to the
similar ionic radii of La and Ba \cite{akahoshi}.
We therefore set $\delta \phi = \phi^{La}_0 - \phi_0$, where 
$\phi^{La}_0 = 175^{\circ} $ is the average bond angle in La$_{0.5}$Ba$_{0.5}$MnO$_3$
\cite{chmaissem}. $\phi_0$ is assumed to decrease from $175^{\circ}$ to $164^{\circ}$, as R changes from La to Y.
Our parameter choice, $\lambda =1.4$, $U'=6$ and $J_S = 0.08$ is guided by
the observed scales for the N\'eel temperature 
in CaMnO$_3$ and the transport gap in LaMnO$_3$ \cite{jirak,palstra}. 

Certain limits of the model Hamiltonian Eq. (1) have been analyzed already.
The effects of large $U'$ have been addressed within mean-field theories
for a restricted choice of magnetic phases \cite{maezono,mishra}.
In the absence of the Hubbard term, various ordered phases were discovered
at half-doping using an exact diagonalization (ED) based Monte-Carlo method \cite{dag_x0.5}.
The effect of on-site disorder near a first-order phase boundary was reported recently \cite{sen_PRL}.
But a realistic and simultaneous treatment of bond- and potential-disorder
and the explicit treatment of an inter-orbital Hubbard term,
while exactly retaining the spatial correlations has so far been lacking.

Here, we use a combination of the unrestricted Hartree-Fock (HF) scheme and the travelling cluster approximation (TCA).
The TCA allows for sampling of classical configurations for spin and lattice variables according to the Boltzmann weight, and involves
iterative ED of small clusters \cite{TCA}. The Hartree-Fock decomposition of the Hubbard term leads to three independent HF
parameters per site $\langle n_{ia} \rangle$, $\langle n_{ib} \rangle$ and 
$\langle \tau^{+}_{i} \rangle = \langle c^{\dagger}_{i b} c^{}_{ia} \rangle$, which enter as additional parameters in TCA.
These HF parameters are self-consistently evaluated with the annealing process of the TCA.
Most results are on lattices with $N=16^2$ sites using a travelling cluster 
with $N_c = 4^2$.
\begin{figure}[t!]
\centerline{\includegraphics[width=9cm]{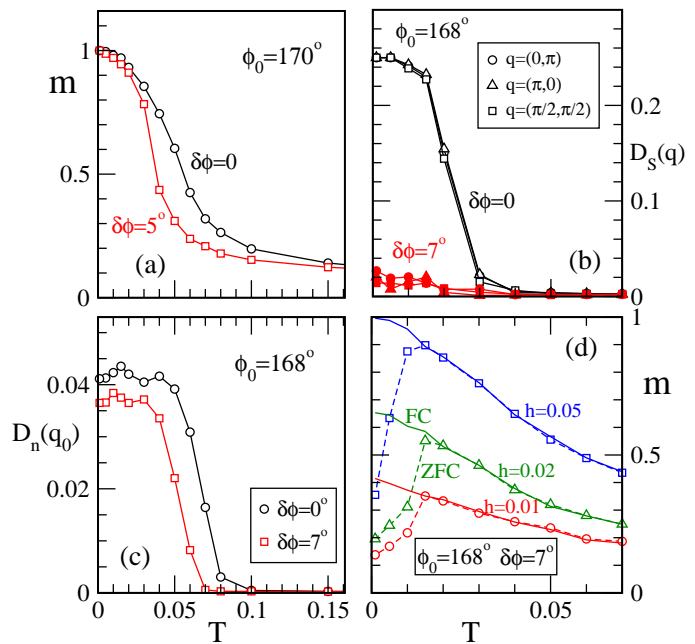}}
\caption{(Color online)~ Temperature dependence of (a) the magnetization,
(b) the spin structure factor at selected momenta, and
(c) the staggered charge structure factor
for clean and bond-disordered systems. 
(d) Field-cooled (FC) and zero-FC magnetization
for the bond-disordered system.
}
\end{figure}

The indicators for spin- and charge ordering are plotted in Fig. 1. 
Panel (a) shows the temperature dependent
magnetization, defined via $m^2 = \langle ( N^{-1} \sum {\bf S}_i )^2 \rangle_{av}$,
for clean and bond-disordered systems for 
$\phi_0 = 170^{\circ}$. Here and below
$\langle ... \rangle_{av}$ denotes 
the average over thermal equilibrium configurations, 
and additionally over realizations of quenched disorder.
Bond disorder with $\delta \phi = 5^{\circ}$ considerably reduces the
Curie temperature $T_C$, which is estimated from the inflection-point in $m(T)$.
Recall that the strength of bond-disorder $\delta \phi$ is tied to the average
bond angle $\phi_0$ via $\delta \phi = \phi^{La}_0 - \phi_0$.
In the clean system $m(T)$ and in particular $T_C$, are not affected
much upon varying $\phi_0$.
The fingerprint of CE spin order is the simultaneous presence of peaks
at wavevectors ${\bf q} = (0,\pi)$, $(\pi,0)$ and $(\pi/2,\pi/2)$, in the spin structure factor
$D_S({\bf q}) = N^{-2} \sum_{ij} \langle {\bf S}_i \cdot {\bf S}_j \rangle_{av} ~ 
e^{-{\rm i} {\bf q} \cdot ({\bf r}_i - {\bf r}_j)}$. In Fig. 1(b), a simultaneous rise in the
three components of $D_S({\bf q})$ is observed for $\delta \phi = 0^{\circ}$, whereas in the 
disordered case no sizable peaks appear at any ${\bf q}$. 

For the clean system, the charge structure factor $D_n({\bf q}) = N^{-2} \sum_{ij} \langle n_i \cdot n_j \rangle_{av} ~ 
e^{-{\rm i} {\bf q} \cdot ({\bf r}_i - {\bf r}_j)}$ at ${\bf q} = {\bf q}_0 \equiv (\pi,\pi)$ rises sharply
upon cooling (see Fig. 1(c)), indicating the onset of
staggered charge order. The temperature scales $T_{CO}$ and $T_{CE}$ are inferred from
inflection points in the $T$-dependence of the relevant components of the charge and spin structure factors. 
Charge ordering is accompanied by the ordering of lattice and orbital variables.
The $U'$ term enhances $T_{CO}$, which is otherwise of the
same order as $T_{CE}$ for the values of $\lambda$ used here \cite{brey_x0.5}.
Surprisingly, a clean signal for the onset of charge ordering near $T \sim 0.06$ is found
even in the bond-disordered system. This highlights the crucial qualitative difference
between the effects of bond and 
on-site disorder. The latter is known to strongly suppress charge ordering tendencies \cite {motome}.
Upon comparing Fig. 1(b) and Fig. 1(c) it becomes clear that the
bond-disorder leads to a state which is ordered in the orbital and the charge sector but disordered
in the spin sector. The $U'$ term further stabilizes the charge order in this state.

Additional information about the spin state that emerges in
the presence of bond disorder is obtained
from the field-cooled (FC) and zero-field-cooled (ZFC) magnetizations,
which we calculate by including a Zeeman term $-h\sum_i S^z_i$, in the Hamiltonian.
While the FC and ZFC magnetizations are indistinguishable for clean system, their difference at low $T$
may serve as an
indicator for a spin-glass character for bond-disordered system (see Fig. 1(d)) \cite {SG-comment}.
The temperature for which the FC and ZFC results begin to differ, provides
an estimate for the spin-glass crossover temperature $T_g$.
Such a state with charge and orbital order but glassiness in the spin sector was
recently reported for single layered manganites \cite {mathieu-EPL}.

Fig. 2(a) shows the phase diagram in the $T-\phi_0$ plane, for the clean system. For
$\phi_0 \geq 170^{\circ} $, the system undergoes a paramagnet (PM) to ferromagnet transition upon cooling.
For $\phi_0 \leq 169^{\circ} $, charge and orbital degrees of freedom order at low temperatures,
followed by a transition from a PM to a CE spin state near $T \sim 0.02$. 
The $\phi_0$-driven transition at low $T$ results from the reduction in bandwidth
upon decreasing $\phi_0$. Thereby the effective $\lambda$ and $J_S$ are enhanced leading to
a first-order transition towards a charge and orbital ordered (CO-OO) CE state by opening a gap in the spectrum,
at a critical value $\phi_0^C$.
$\phi_0^C$ increases from $155^{\circ}$ for $U'=0$, to $169^{\circ}$ for $U'=6$.
\begin{figure}[t!]
\centerline{\includegraphics[width=8.6cm, clip=true]{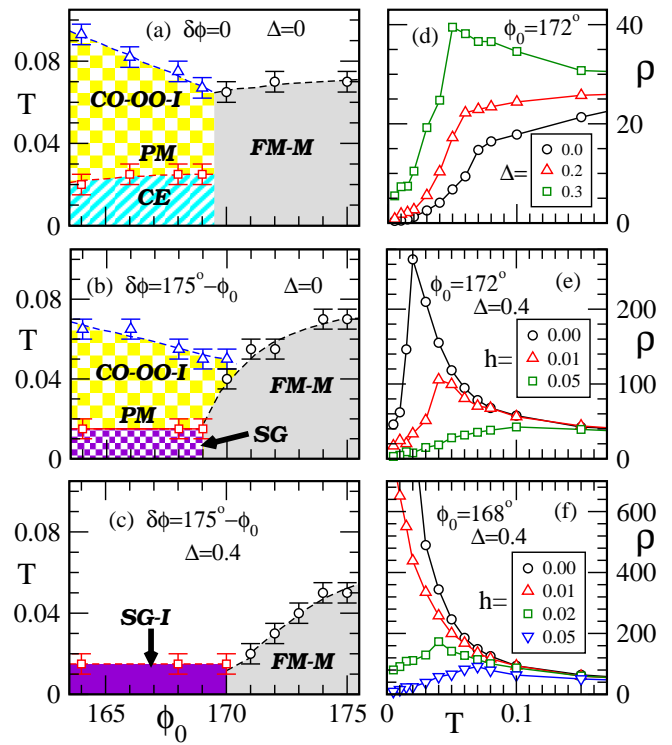}}
\caption{(Color online)~ $T-\phi_0$ phase diagrams at 
$\lambda=1.4$, $J_S = 0.08$, and $U' = 6$, 
(a) $\delta \phi = 0^{\circ}$, $\Delta = 0$,
(b) $\delta \phi = 175^{\circ}-\phi_0$, $\Delta = 0$, and (c) $\delta \phi = 175^{\circ}-\phi_0$, $\Delta = 0.4$.
Symbols are data points and dotted lines are guides to eye.
FM (PM) denotes a ferromagnetic (paramagnetic) state; M (I) indicates metallic (insulating) character;
CO (OO) refers to charge (orbital) order and SG marks the spin-glass state. 
(d) $T$-dependent resistivity $\rho$ for increasing strength of on-site
disorder, in units of $\hbar /\pi e^2$. $\rho(T)$ for varying magnetic field for
$\Delta=0.4$, and (e) $\phi_0 = 172^{\circ}$, (f) $\phi_0=168^{\circ}$.
}
\end{figure}
While charge and orbital order is only weakly affected by bond disorder, the
spin degree of freedom reacts more sensitively (see Fig. 2(b)). For $\phi_0 \leq 169^{\circ} $, glassiness in the spin sector
is induced (see Fig. 1(d)), and for $172^{\circ} \geq \phi_0 \geq 170^{\circ} $, $T_C$ is strongly suppressed.

Some of the key features of the experimental phase diagram for the "disordered" materials
are not captured by bond disorder alone, i.e., there is no charge disordered SG state
and a FM-M to PM-I transition exists only in a very narrow $\phi_0$
window. Therefore, we include additional on-site disorder.
Charge order is lost for finite on-site disorder,
while the spin-glass phase persists (see Fig. 2(c)).
The ground state becomes an unsaturated FM for $170^{\circ} > \phi_0 > 173^{\circ}$ and $T_C$ is further reduced.
The phase diagram in Fig. 2(c) compares very well with the experimental phase diagram of
half-doped manganites \cite{akahoshi}.

Fig. 2(d) shows the effect of on-site disorder on the resistivity $\rho$ of a bond-disordered
system with a ferromagnetic metallic ground state. $\rho$ is approximated by the inverse of 
$\sigma(\omega_{min})$, where $\omega_{min}=10 t/N \sim 0.04 t$ is the lowest reliable energy scale 
for calculations of the optical conductivity $\sigma(\omega)$ on our $16^2$ system \cite {transport}.
Insulating or metallic character is determined from the sign of the slope of $\rho(T)$.
Increasing the strength $\Delta$ of the on-site disorder leads to a reduction in
$T_C$ and an increase in the resistivity, and an FM-M to PM-I transition is observed for $\Delta = 0.3$.

If the ground state is ferromagnetic, the resistivity in an external field drops near
$T_C$ (see Fig. 2(e)), similar to the experiments in Nd$_{0.5}$Ba$_{0.5}$MnO$_3$.
A magnetic field reduces the resistivity in the SG phase at low $T$;
additionally $d\rho/dT$ changes sign, indicating an insulator to metal transition
(see Fig. 2(f)).
Such transitions are indeed observed in (Sm$_{0.3}$Gd$_{0.7}$)$_{0.55}$Sr$_{0.45}$MnO$_3$
\cite{tomioka, maignan}.
\begin{figure}[t!]
\centerline{\includegraphics[width=9cm, clip=true]{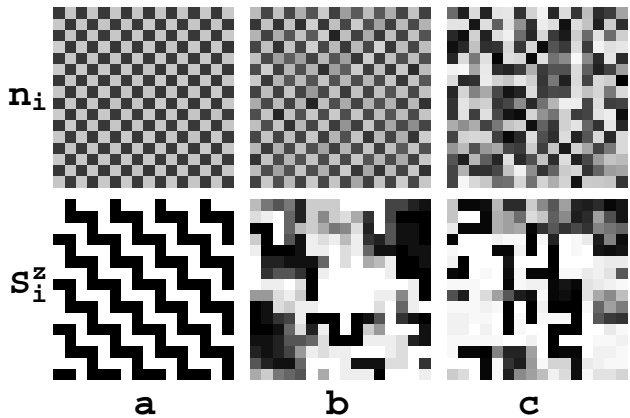}}
\caption{~ Monte Carlo snap-shots for $\phi_0 = 168^{\circ}$ on a $16 \times 16$ lattice.
Column-(a): $\delta \phi = 0^{\circ}$, $\Delta=0$, column-(b): $\delta \phi = 7^{\circ}$, $\Delta=0$ and
column-(c): $\delta \phi = 0^{\circ}$, $\Delta=0.4$. The top row shows the
charge density $n_i$, grayscale covering the range from 0.2(white) to 0.8 (black). 
The bottom row plots the spin component $S_i^z$, grayscale from -1 (white) to 1 (black).
Results are shown for specific realizations of disorder at $T=0.005$.}
\end{figure}

Fig. 3 shows real-space patterns for the charge $n_i$ and the spin $S^z_i$
variables for clean (column a), bond disordered (column b), and on-site disordered (column c) 
systems. 
The groundstate in the clean case is a CE phase with charge and orbital order.
A checkerboard pattern for the charge density
and a zig-zag FM chain structure for spin variables characterize this phase \cite {dag_x0.5}.
Bond disorder does not affect the charge ordering, while the spins ${\bf S}$ become
disordered leading to a nontrivial state with simultaneous charge order and spin glassiness.
On-site disorder spoils the charge ordering and the magnetic order.
The spatial patterns reveal inhomogeneities on
the scale of a few lattice spacings.
Since charge order can exist in the PM phase at elevated temperatures (see Fig. 2(a), (b)), 
its coexistence with spin-glass behavior is not surprising. The different effects of
bond disorder on charge and spin degrees of freedom is, however, non-trivial.
On the one hand, a decreasing bond-angle leads to a lowering of the fermionic kinetic
energy across that bond and thereby reduces the tendency towards double exchange ferromagnetism. But more importantly,
the zig-zag structure of the spin-aligned chains in the CE phase is disrupted by random weak bonds.
The charge variables instead vary only weakly with bond disorder,
which explains the stability of the charge ordered state. 

In summary, a two-band double-exchange model with electron-lattice and
electron-electron interactions with bond and on-site disorder
provides the basis for 
an overall understanding of the experimental 
phase diagram of R$_{0.5}$Ba$_{0.5}$MnO$_3$ \cite{akahoshi}.
Our results reveal the existence of a non-trivial phase, which is
ordered in the charge variables but glassy in the spin variables.
The bond disorder is important for the spin-glass behavior, 
and likely to explain the coexistence of charge order and spin glassiness in
single-layered manganites \cite {mathieu-EPL}.
On-site disorder is crucial for a description of the thermally driven metal to insulator
transitions.
Each of the two types of disorder explain selected features in manganites, but only a combination
of both can describe most of the experimentally observed phenomena.

We acknowledge support by the 
Deutsche Forschungsgemeinschaft through SFB 484, and the use of
the Beowulf Cluster at HRI, Allahabad (India).

{}


\begin{thebibliography}{}
\bibitem{SG-comment} The confirmation of a spin-glass phase requires access to the spin dynamics.
This however is beyond the scope of the present study, and we therefore take the difference
between FC and ZFC magnetizations as our working definition
for the spin glass state.
\end{thebibliography}
\end{document}